\documentclass[aps,twocolumn,gbroupedaddress,amsmath,amssymb]{revtex4-2}
\usepackage[utf8]{inputenc}
\usepackage[T1]{fontenc}
\usepackage{amsmath}
\usepackage{amsfonts}
\usepackage{amssymb}
\usepackage{graphics,graphicx}
\usepackage{graphicx}
\usepackage{subfigure} 
\usepackage{wrapfig} 
\usepackage{epstopdf}
\usepackage{xcolor}
\graphicspath{{figuras/}}
\def\be{\begin{equation}}
\def\ee{\end{equation}}
\def\ba{\begin{eqnarray}}
\def\ea{\end{eqnarray}}

\usepackage{ulem}
\usepackage[breaklinks=true]{hyperref}
\usepackage{setspace}
\usepackage{graphicx}
\usepackage{color}

\begin{document}
	
\title{Effective four-dimensional loop quantum black hole with a cosmological constant}
    \author{Jianhui Lin}
\affiliation{Department of Physics, South China University of Technology, Guangzhou 510641, China}
	\author{Xiangdong Zhang}\email{Corresponding author. scxdzhang@scut.edu.cn}
\affiliation{Department of Physics, South China University of Technology, Guangzhou 510641, China}

	\begin{abstract}
	In this paper, we utilize the effective corrections of the $\bar{\mu}$-scheme in loop quantum black holes to obtain a 4-dimensional spherically symmetric metric with a cosmological constant. By imposing the areal gauge on the components of Ashtekar variables in the classical theory and applying the holonomy corrections, we derive the equations of motion, which can be solved to obtain the expression for the effective metric in the Painlev\'{e}-Gullstrand coordinates. Compared to the classical dS (AdS) spacetime, the LQG correction sets an upper bound on the cosmological constant as $\Lambda<\frac{3}{\gamma^2\Delta}$. The thermodynamic properties of black holes have also been calculated. We interestingly found that for a small black hole, the temperature of the LQG black hole decreases as the mass decreases, which is quite different with the classical scenario. Moreover, our result shows that a logarithmic term appeared as the leading order correction to the Beikenstein-Hawking entropy. Furthermore, the LQG corrections also introduce an extra phase transition in the black hole's heat capacity at smaller radius. 
	\end{abstract}
	\maketitle
\section{Introduction}

The prediction of the singularity at the center of black holes by classical General Relativity (GR) has led to the widespread belief that classical theory has limitations and that quantum gravity effects need to be introduced to cure this type of spacetime singularity. Pursuing a consistent quantum gravity theory becames one of the greatest challenges since the 20th century.

 One approach to investigating how quantum gravity affects the spacetime of black holes is to start with a specific quantum gravity theory and determine its model corresponding to spherically symmetric spacetime, and then make physical predictions based on the model. loop quantum gravity (LQG) is currently one of the candidates for a theory of quantum gravity. As a background independent and nonperturbative theory, it has considerable appeal in this regard (see, e.g., \cite{Ashtekar 2004, Rovelli 2004, Thiemann, Han 2007}). Since the late 1980s, LQG based on Ashtekar variables has seen significant development up to the present day. This includes natural predictions for discrete geometrical spectrum \cite{Rovelli 1995, Ashtekar 1997-01, Ashtekar 1997-11, Yang 2016, Thiemann 1998, Ma 2010} and successful generalizations to metric $f(R)$ theories, higher-dimensional gravity and so on \cite{Zhang 2011, Bodendorfer 2013, Zhang 2020}. It preserves two fundamental principles of general relativity: diffeomorphism invariance and background independence. The construction of LQG adheres to mathematical rigor and physical self-consistency. In situations where current experimental conditions cannot be satisfied, it is necessary to theoretically consider whether the classical limit of quantum theory is correct. 
 
 The application of LQG to cosmology, known as loop quantum cosmology (LQC), has been established as an appropriate semiclassical state \cite{Tan 2006}, and the expectation values of the quantum operators in this state match well with their corresponding classical values. It has also led to the conclusion that the "Big Bounce" replaces the big bang, successfully resolving the problem of the cosmological singularity of the big bang \cite{Bojowald 2001, Modesto 2004, Ashtekar and Bojowald 2005}. LQC is the symmetry-reduced model of LQG \cite{Ashtekar 2003}. In the classical scenario, before quantization, we can utilize the symmetries of spatial homogeneity and isotropy to reduce the phase space of gravitational degrees of freedom from infinite dimensions to finite dimensions. Then, by using the methods and techniques of LQG, it is able to proceed with its quantization. For a detailed review of LQC, see, e.g., \cite{Bojowald 2008, Ashtekar 2011}. 
 
 Due to the success of LQC in resolving classical cosmological puzzles, the attempt to apply this technique to black holes to address the issue of black hole singularity is a very intriguing idea. As the simplest black hole solution, the Schwarzschild black hole then serves as an ideal arena to implement these ideas. Note that the interior of Schwarzschild black hole is isometric to the Kantowski-Sachs model \cite{Ashtekar and Bojowald 2005, Boehmer 2007}, thus allowing for the potential adaptation of techniques and ideas from LQC. Therefore, in the past decade or so, the exploration of loop quantum black hole models has been a popular direction \cite{Ashtekar and Bojowald 2005,Boehmer 2007, D.-W. 2008,Ashtekar 2018, Zhang2020}. 
 
Since the success of LQC is based on its unique $\bar{\mu}$ quantization scheme. It is therefore great interest to also implement the $\bar{\mu}$ scheme in loop quantum black hole models. In the quantum effective Hamiltonian constraint, the holonomy correction is simplified by replacing the components of the Ashtekar connection with quantum corrections, which are controlled by the quantum regularization parameters due to the fundamental discreteness of LQG.  In the $\bar \mu$ scheme, the quantum parameters are chosen as adaptive discreteness variables. However, the $\bar{\mu}$-scheme encounters an issue of excessive quantum corrections at the event horizon \cite{Zhang 2023}. In the classical regime, excessive quantum corrections are generally considered unacceptable. Moreover, in \cite{Sartini 2021}, the unimodular formulation of general
 relativity is proposed to study the full quantum dynamics of the LQC and the unimodular representation is applied to the homogeneous black hole interior spacetime
 such that we can test the various regularization schemes. It is showed that in $\bar{\mu}$ scheme, the area of the $2-$sphere is smaller than the area gap in LQG, which violates the very construction of the $\bar{\mu}$ scheme. Fortunately, recently, as suggested by some authors \cite{Ewing 2020}, the reason of traditional $\bar{\mu}$-scheme cannot be implemented is simply because we choose the wrong set of coordinates which becomes
null at the horizon. Hence they suggested to implement the $\bar{\mu}$-scheme in terms of another set of coordinates which will not become null at the horizon. The use of everywhere spacelike Painlev\'{e}-Gullstrand coordinates can be employed for the $\bar{\mu}$-scheme and avoid the aforementioned problem, leading to an effective framework for spherically symmetric vacuum solutions with holonomy corrections from loop quantum gravity \cite{Ewing 2020}. 


On the other hand, up to now, most of the studies onT loop quantum black hole models are limited to the Schwarzschild case. Note that our current universe is undergoing an accelerating expansion and leading to the famous dark energy issue. The origin of dark energy remains as one of the biggest challenges to modern physics. Many possible mechanisms have been
proposed to account for this issue, such as the phenomenological models\cite{Friemann08}, modified gravity \cite{Banerjee,Sen,Qiang,Peebles03}, higher dimensions \cite{Tye01} and so on. Among them, the cosmological constant is generally
believed as the most simplest explanation\cite{Peebles03,Weinberg89}. Though people may argue that the observed cosmological constant $\Lambda\approx 10^{-52}m^{-2}$ is so small and hence its effects can be safely ignored. However, the existence of a cosmological constant, regardless its value, will dramatically changed the asymptotic structure of the spacetime \cite{Ashtekar16}. In addition, inspired by AdS/CFT correspondence, the black hole solutions with a negative cosmological constant also should be considered.  
Moreover, in Schwarzschild-AdS case, we have more richer physics such as
Hawking-Page phase transition and the extended phase space
thermodynamics can be established. Hence, the inclusion of a cosmological constant in a black hole solution is important both in the practical and the theoretical sense. 

Additionally, thermodynamic properties of black holes have been studied for many years.  Corresponding to the standard laws of thermodynamics, black hole thermodynamics also has four laws. Just like the standard laws of thermodynamics, black hole thermodynamics also follows four fundamental laws \cite{Bardeen 1973}. The first law, for example, establishes an energy conservation equation that connects the entropy of a black hole to its mass, charge, rotation, and other parameters \cite{Wald 2001}. As a thermodynamic system, black holes can be assessed for their thermodynamic stability. For instance, the Schwarzschild black hole, has a negative heat capacity, indicating that it is thermodynamically unstable. Also, it was discovered that the spacetime of black holes can exhibits a rich phase structure and critical phenomena (see e.g. \cite{Kubiznak 2012} \cite{Dayyani 2018} \cite{Wei 2020}), which are entirely similar to those found in other known thermodynamic systems. Within this context, if we aim to explore whether black holes exhibit phase transition behavior similar to that of a van der Waals fluid, we need to extend the metric to include a cosmological constant. This is because, in the extended phase space, the pressure is linked to the cosmological constant, which is no longer treated as a constant \cite{Kubiznak 2012}. Therefore, to calculate the possible phase transition behavior of black holes in LQG, we must extend the effective metric to include a cosmological constant. Therefore, in this paper, we will follow the line of \cite{Ewing 2020} by utilizing $\bar{\mu}$ scheme in Painlev\'{e}-Gullstrand coordinates to study the loop quantization of Schwarzschild de-Sitter(anti de-Sitter) black holes.

 This paper will be organized as follows: We will start from the classical Hamiltonian framework and, in Section \ref{Sec-Classical theory}, present the specific form of the metric using the components of connection and densitized triad as dynamical variables. In Section \ref{Sec-Effective correction of LQG}, the LQG holonomy corrections are employed and obtain corresponding metric solutions. Additionally, a discussion of the physical properties of the solutions is provided in Section \ref{Physicalpropertiesofmetric}. Having obtained the effective metric with a cosmological constant, in Section \ref{Thermodynamic LQG}, we investigate the phase transitions behavior of AdS black holes in LQG. Finally, Section \ref{Sec-Conclusions and discussions} concludes the article with a summary.
 
 Our conventions are the following: space-time indices are denoted by $\mu, \nu, \rho,\sigma,...$; spatial indices are denoted by $a; b; c; ...$; and internal indices are denoted by $i, j, k,...$ We only set $c=1$ but keep $G$ and $\hbar$ explicit.

	\section{CLASSICAL THEORY}\label{Sec-Classical theory}
	
	The metric of a 4-dimensional spherically symmetric spacetime can be expressed in Painlev\'{e}-Gullstrand coordinates as
	
	\begin{eqnarray}\label{metric}
		ds^2=-N^2dt^2+f^2(dx+N^xdt)^2+y^2d\Omega^2,
	\end{eqnarray} 
	 where the lapse $N$, shift vector $N^x$ and $f,y$ all are the function of time $t$ and the radial coordinate $x$, while $d\Omega$, given by $d\Omega^2=d\theta^2+\sin^2\theta d\phi^2$, is the line element on the unit sphere. 
	 
	 As the foundation of LQG, connection dynamics is a theory based on the Hamiltonian formulation of GR, described by Ashtekar-Barbero connections and their conjugate momentum in terms of triads. In loop quantum black hole models, we also inherit this feature and use them as basic variables. In this context, the basic variables are consistent with \cite{Ewing 2020}, which respectively are the Ashtekar-Barbero connection components given by $c(x)$, $p(x)$, and the densitized triad components given by $E^c(x)$, $E^p(x)$. 
	 The following represent the densitized triads in terms of metric components \cite{Ewing 2020}
	 \begin{eqnarray}\label{triads}
	 	E^x_1&=&y^2sin\theta=E^csin\theta,  \qquad E^\theta_2=fysin\theta=E^psin\theta,\nonumber \\ 
	 	E^\phi_3&=&fy=E^p.
	 \end{eqnarray}
	  Moreover, the Ashtekar-Barbero connection are written as $A^i_a=\Gamma^i_a+\gamma K^i_a$, where both the spin-connection $\Gamma^i_a$ and the extrinsic curvature $K^i_a$ only have three non-zero components \cite{Ewing 2020}
	 \begin{align}
	 	K^1_x&=\frac{c}{\gamma}, &\quad K^2_\theta&=\frac{p}{\gamma}, &\quad K^3_\phi&=\frac{psin\theta}{\gamma} \nonumber \\
	 	\Gamma^3_\theta&=-\frac{\partial_x E^c}{2 E^p}, &\quad \Gamma^1_\phi&=-cos\theta, &\quad \Gamma^2_\phi&=\frac{\partial_xE^csin\theta}{2E^p},
	 \end{align} 
	 where $\gamma$ is the Barbero-Immirzi parameter (a dimensionless constant that labels various inequivalent kinematic quantizations of LQG). The determinant of the spatial metric is $q=E^c (E^psin\theta)^2$.
	 
     \subsection{Actions and constraints}
	 
	 The action of GR plus a cosmological constant reads
	 \begin{eqnarray}
	 	S=\frac{1}{16 \pi G} \int d^4x \sqrt{-g} \left(R-2\Lambda\right).
	 \end{eqnarray}
	 After performing canonical analysis, the gravitational action becames
	 \begin{eqnarray}
	 	S=\int dt\int_{\Sigma}\left[\frac{\dot{A}^i_a E^a_i}{8\pi G\gamma}-N\mathcal{H}_\Lambda-N^a\mathcal{H}_a\right],
	 \end{eqnarray}
	where the dot means the derivative with respect to time $t$; the Hamiltonian constraint $\mathcal{H}_\Lambda$ and the diffeomorphism constraintis reads respectively as
	 \begin{eqnarray}
	 	\mathcal{H}_\Lambda&=&-\frac{E^a_iE^b_j}{16\pi G\gamma ^2\sqrt{q}}{\epsilon^{ij}}_k({F_{ab}}^k-(1+\gamma^2){\Omega_{ab}}^k)) \nonumber \\
	 	&&-\frac{\sqrt{q}N}{8\pi G}\Lambda.\\
	 	\mathcal{H}_a&=&\frac{1}{8\pi G\gamma}E^b_k {F_{ab}}^k.
	 \end{eqnarray}
	 Here the field strength is given by $F_{ab}{}^{k}=2\partial_{[a}A_{b]}^{k}+\epsilon_{ij}{}^{k}A_{a}^{i}A_{b}^{j}$ and the spatial curvature denotes $\Omega_{ab}{}^k=2\partial_{[a}\Gamma_{b]}^k+\epsilon_{ij}{}^k\Gamma_{a}^i\Gamma_{b}^j$.
	 Due to the spherical symmetry, we can integrate over $d\Omega$ and use the Eq. \eqref{triads}, then obtain the following expression
	 \begin{eqnarray}\label{action}
	 	S=\int dt\int dx{\left[\frac{\dot{c}E^c+2\dot{p}E^p}{2G\gamma}-N\mathcal{H}_\Lambda-N^x\mathcal{H}_x\right]},
	 \end{eqnarray}
	 where the Hamiltonian constraint reads
	 \begin{eqnarray}
        \mathcal{H}_\Lambda = &-&\frac{1}{2G\gamma}\left[\frac{2cp\sqrt{E^c}}{\gamma}+\frac{E^p}{\gamma\sqrt{E^c}}(p^2+\gamma^2)-\frac{\gamma(\partial_xE^c)^2}{4E^p\sqrt{E^c}}\right. \nonumber \\
        &-&\left.\gamma\sqrt{E^c}\partial_x\left(\frac{\partial_xE^c}{E^p}\right)-\gamma \sqrt{E^c}E^p\Lambda\right],
	 \end{eqnarray}
	 and the diffeomorphism constraint is
	 \begin{eqnarray}\label{reduceddiffeomorphism}
	 	\mathcal{H}_x=\frac1{2G\gamma}(2E^p\partial_xp-c\partial_xE^c).
	 \end{eqnarray}
	 Here straightforward calculations show only $\mathcal{H}_x$ is non-zero and $\mathcal{H}_\theta,\mathcal{H}_\phi$ both vanish. Moreover, from Eq. \eqref{action}, the symplectic structure of the symmetry-reduced black hole models is given by
	 \begin{eqnarray}
	 	\{c(x_1),E^c(x_2)\}=2G\gamma\delta(x_1-x_2),
	 \end{eqnarray}
	 \begin{eqnarray}
	 	\{p(x_1),E^p(x_2)\}=G\gamma\delta(x_1-x_2),
	 \end{eqnarray}
	 with $\delta$ being the Dirac delta function. We now consider the smeared constraint function $\mathcal{C}[N]=\int dxN\mathcal{H}_\Lambda$ and $\mathcal{D}[N^x]=\int dxN^x\mathcal{H}_x$. Through lengthy but straightforward calculations, the Poission brackets between constraints algebra read
	 \begin{eqnarray}
	 	\{\mathcal{C}[N_1],\mathcal{C}[N_2]\}=\mathcal{D}\left[\frac{E^c}{(E^p)^2}(N_1\partial_xN_2-N_2\partial_xN_1)\right],
	 \end{eqnarray}
	 \begin{eqnarray}
	 	\{\mathcal{D}[N_1^x],\mathcal{D}[N_2^x]\}=\mathcal{D}[(N_2^x\partial_xN_1^x-N_1^x\partial_xN_2^x)],
	 \end{eqnarray}
	 \begin{eqnarray}
	 	\{\mathcal{C}[N],\mathcal{D}[N^x]\}=-\mathcal{C}[N^x\partial_xN].
	 \end{eqnarray}
	 As compared to the case without the cosmological constant \cite{Ewing 2020}, the symplectic structure and the constraint algebra have not changed. Using the Hamilton evolution equations $\dot{y}=\left\{y,\mathcal{C}[N]+\mathcal{D}[N^x]\right\}$, the equations of motion for each component with respect to time are determined by
	 \begin{eqnarray}
	 	\dot{E}^c=\frac{2Np}\gamma\sqrt{E^c}+N^x\partial_xE^c,
	 \end{eqnarray}
	 \begin{eqnarray}\label{Ep}
	 	\dot{E}^p=\frac N{\gamma\sqrt{E^c}}(cE^c+pE^p)+\partial_x(N^xE^p),
	 \end{eqnarray}
	 \begin{align}
	 	\dot{c}=&\frac N{2\gamma\sqrt{E^c}}\left[\frac{E^p}{E^c}(p^2+\gamma^2)-2cp\right] +\gamma\partial_x\biggl(\frac{\partial_x(N\sqrt{E^c})}{E^p}\biggr)  \nonumber \\
	 	&+\frac{N\gamma}{2\sqrt{E^c}}\left[\partial_x\biggl(\frac{\partial_xE^c}{E^p}\biggr)-\frac{(\partial_xE^c)^2}{4E^cE^p}\right]+\partial_x(N^xc)\nonumber \\ &-\frac\gamma2\partial_x{\left(N\frac{\partial_xE^c}{E^p\sqrt{E^c}}\right)}+\frac{N\gamma E^p\Lambda}{2\sqrt{E^c}}
	 \end{align}
	 \begin{eqnarray}\label{p}
        \dot{p}=&-&\frac{N}{2\gamma\sqrt{E^c}}(p^2+\gamma^2)-\frac{\gamma}{2}\left[\frac{N}{\sqrt{E^c}}\left(\frac{\partial_xE^c}{2E^p}\right)^2\right.\nonumber \\&-&\left.\partial_x(N\sqrt{E^c})\frac{\partial_xE^c}{(E^p)^2}+N^x\partial_xc \right]+\frac{N\gamma \sqrt{E^c}\Lambda}{2}
	 \end{eqnarray}
	Once the lapse and shift vector are chosen, by solving the above equations of motion and the diffeomorphism constraint together with the Hamiltonian constraint, one can obtain the vacuum spherically symmetric metric with the cosmological constant. 
	
	\subsection{Classical solutions}
	Following the same approach in \cite{Ewing 2020}, we also adapt the areal gauge for $E^c$ as follows:
	\begin{eqnarray}\label{areal gauge}
		E^c=x^2.
	\end{eqnarray}
	Once the areal gauge is imposed, the diffeomorphism constraint \eqref{reduceddiffeomorphism} then in turn implies \ba
\quad c=\frac{E^p}{x}\partial_x p.
\ea The gauge-fixing condition $\zeta=E^c-x^2$ is second class with the diffeomorphism constraint $\mathcal{H}_x$, which should be preserved by the equations of motion. That is to say, $\dot{\zeta}=0$. Using the condition $\dot{E}^c=0$, we can obtain the expression 
	\begin{eqnarray}\label{relation about N^x and N}
		N^x=-\frac{Np}{\gamma},
	\end{eqnarray}
	which implies that $N^x$ is no longer a Lagrange multiplier. After choosing the areal gauge, $N^x$ is determined by the lapse function $N$. Through this process, the action can be simplified and given by:
	\begin{eqnarray}\label{simplified action}
		S_{GF}=\int dt \int dx\left[\frac{\dot{p}E^p}{G\gamma}-N\mathcal{H}_\Lambda\right]
	\end{eqnarray}
	with
	\begin{align}\label{effective Hamiltonian}
		\mathcal{H}_\Lambda=&\frac1{2G\gamma}\left[\frac{3\gamma x}{E^p}-\frac{2\gamma x^2}{(E^p)^2}\partial_xE^p-\frac{E^p}{\gamma x}\partial_x[x(p^2+\gamma^2)]\right]\nonumber\\
		&+\frac{xE^p}{2G}\Lambda.
	\end{align}
	Since the $E^c$ is fixed by the areal gauge, the remaining symplectic structure of the symmetry-reduced theory reads
	\begin{eqnarray}
		\{p(x_1),E^p(x_2)\}=G\gamma\delta(x_1-x_2),
	\end{eqnarray}
	and similarly, the remaining constraint algebra is
	\begin{eqnarray}\label{constraint algebra}
		\{\mathcal{C}[N_1],\mathcal{C}[N_2]\} &=&C{\left[-\frac1\gamma\left(N_1\partial_xN_2-N_2\partial_xN_1\right)p\right]}  \nonumber\\
		&=&C\left[N_1^x\partial_xN_2-N_2^x\partial_xN_1\right],
	\end{eqnarray}
	where the second step utilizes Eq.\eqref{relation about N^x and N}, which corresponds to the Painlev\'{e}-Gullstrand coordinates. The equations of motion can also be simplified to
	\begin{align}
		\dot{E}^p=&\frac{p}{\gamma x}\left(NE^p-x\partial_x\left(NE^p\right)\right),\\
		\dot{p}=&\frac{\gamma Nx}{2(E^p)^2}+\frac{\gamma x^2}{(E^p)^2}\partial_xN-\frac{N}{2x\gamma}\partial_x\left(xp^2+\gamma^2x\right)\nonumber\\
		&+\frac{\gamma Nx}{2}\Lambda,
	\end{align}
	 which can be obtained by substituting areal gauge \ref{areal gauge} and the expression \eqref{relation about N^x and N} into the equations of motion \eqref{Ep} and \eqref{p}. Now we set $N=1$, and take into account the Hamiltonian constraint $\mathcal{H}_\Lambda=0$. Then, $\dot{E}^p=0$ and $\dot{p}=0$ can be solved to obtain 
	 \begin{eqnarray}
	 	E^p&=&C_1x,\label{classical solution1}\\
	 	p^2&=&\gamma^2\left(\frac{1}{{C_1^2}}-1\right)+\frac{\gamma^2x^2\Lambda}{3}+\frac{C_2}{x},\label{classical solution2}
	 \end{eqnarray}
	 where $C_1$ and $C_2$ are two constants of integration. Furthermore, these solutions \eqref{classical solution1} and \eqref{classical solution2} satisfy the Hamiltonian constraint equation $\mathcal{H}_\Lambda=0$ identically.
     From equation \eqref{triads}, it is worthy noting that both $f$ and $y$ can be expressed in terms of $E^c$ and $E^p$ with 
     \begin{eqnarray}
     	f^2&=&\frac{(E^p)^2}{E^c},\label{f square}\\
     	y^2&=&E^c.\label{y square}
     \end{eqnarray}
      Using the equations \eqref{relation about N^x and N}, \eqref{f square}, \eqref{y square} and $N=1$, by inserting the solutions \eqref{classical solution1} and \eqref{classical solution2} into the  metric \eqref{metric}, we obtain
     \begin{eqnarray}\label{PG metric}
     	ds^2=&-&\left(1-\frac{2GM}{x}-\frac{x^2\Lambda}{3}\right)dt^2\nonumber\\&+&2\sqrt{\frac{2GM}{x}+\frac{x^2\Lambda}{3}}dxdt
     	+dx^2+x^2d\Omega^2,
     \end{eqnarray}
      where we can determine the integration constants $C_1=1$ and $C_2=2GM\gamma^2$ by comparing with the standard Schwarzschild de-Sitter(anti-de Sitter) solution and $M$ is the mass of the black hole. Moreover, we can also confirm that the square root of $p$ is negative. Finally, the solutions to the equations of motion are given by
      \begin{eqnarray}
      	E^p&=&x,\\
      	p&=&-\sqrt{\frac{2GM\gamma^2}{x}+\frac{\gamma^2x^2\Lambda}{3}}.\label{classical solution of p}
      \end{eqnarray}
     
     \section{Effective correction of LQG}\label{Sec-Effective correction of LQG}
     We adopt the standard LQC treatment for Schwarzschild black holes, which requires to introduce holonomy corrections. The holonomy of an $SU (2)$-connection $A_a^i$ is path-ordering exponential integral along an edge $e^a$ 
     \begin{eqnarray}
     	h(A)=\mathcal{P}exp\int_{e_i}dtA_a^j\tau_je^a,
     \end{eqnarray}
     with $\mathcal{P}$ labeling path-ordering. Although it is not yet fully clear whether the LQG dynamics of a loop quantum Schwarzschild black hole can also provide a good approximation for the full LQG dynamics of the Schwarzschild black hole spacetime, the success of LQC has led people to believe that the effective dynamics of the loop quantum Schwarzschild black hole model can serve as a good approximation for the semiclassical quantum dynamics of observable objects with physical length scales much larger than the Planck scale. As the basic operators in LQG are the holonomies and triads, it is necessary to use the holonomies instead of the Ashtekar-Barbero components in the present paper. Note that we use the areal gauge classically, the components $c$ and $E^c$ is fixed at the classical level. Hence, we only need to replace Ashtekar connection component $p$ with the corresponding holonomy. In $\bar{\mu}$ scheme, this can be realized as $p\rightarrow\frac{1}{\delta_b}\sin\left(\delta_bp\right)$ with $\delta_b=\sqrt{\frac{\Delta}{E^c}}$ \cite{BV07}. In the resulting quantum effective Hamiltonian constraints, the holonomy correction is simplified by replacing the components of Ashtekar connection $p$ with \cite{Ewing 2020}
     \begin{eqnarray}\label{replaced p}
     	p\rightarrow\frac{x}{\sqrt{\Delta}}\sin\left(\frac{\sqrt{\Delta}}{x}p\right)
     \end{eqnarray}
     after the areal gauge is imposed, where $\Delta$ is the smallest nonzero eigenvalue of the area operator \cite{Ashtekar 2004}, satisfying the equation $\Delta=4\sqrt{3}\pi \gamma \ell_{Pl}^2$, and $\ell_{Pl}$ represents the Planck length. The effective Hamiltonian can be constructed by replacing all the instances of $p$ in \eqref{effective Hamiltonian} with \eqref{replaced p} and the result is 
     \begin{eqnarray}\label{H(LQG)}
      \mathcal{H}^{(\mathrm{LQG})}_\Lambda=&-&\frac1{2G\gamma}\left[\frac{E^p}{\gamma x}\partial_x\left(\frac{x^3}\Delta\sin^2\frac{\sqrt{\Delta}p}x+\gamma^2x\right)\nonumber\right.\\&-&\left.\frac{3\gamma x}{E^p}+\frac{2\gamma x^2}{(E^p)^2}\partial_xE^p-x\gamma E^p\Lambda\right].
     \end{eqnarray}
     And we can obtain the Poisson brackets between $\mathcal{C}^{\mathrm{(LQG)}}$
     \begin{align} \label{C(LOG)}
     &\{\mathcal{C}^{\mathrm{(LQG)}}[N_{1}],\mathcal{C}^{\mathrm{(LQG)}}[N_{2}]\} \nonumber\\
     &=C^{(\mathrm{LQG})}\left[-\frac x{\gamma\sqrt{\Delta}}\mathrm{sin}\frac{\sqrt{\Delta}p}x\mathrm{cos}\frac{\sqrt{\Delta}p}x\bigl(N_1\partial_xN_2\nonumber\right.\\&-\left.N_2\partial_xN_1\bigr)\right].
     \end{align}
     The equation \eqref{relation about N^x and N} also needs to be changed, and at the same time, in order to give desired constraint algebra  \eqref{C(LOG)}, The equation \eqref{relation about N^x and N} has been modified to read as
     \begin{eqnarray}\label{NxLQG}
     	N^x=-\frac{Nx}{\gamma\sqrt\Delta}\sin\frac{\sqrt\Delta p}x{\cos\frac{\sqrt\Delta p}x}.
     \end{eqnarray}
	
	Finally just like the classical case, once the effective Hamiltonian \ref{H(LQG)} is in hand, the modified equations of motion can be obtained
	\begin{eqnarray}\label{EpLQG}
		\dot{E}^p=-\frac{x^2}{2\gamma\sqrt\Delta}\partial_x{\left(\frac{NE^p}x\right)}\sin\frac{\sqrt\Delta p}x{\cos\frac{\sqrt\Delta p}x},
	\end{eqnarray}
	\begin{eqnarray}
		\dot{p}&=&\frac{\gamma Nx}{2(E^p)^2}\left(1+2x\frac{\partial_xN}N\right)-\frac{\gamma N}{2x}\nonumber\\&-&\frac N{2\gamma\Delta x}\partial_x\bigg(x^3\sin^2\frac{\sqrt{\Delta}p}x\bigg)+\frac{x\gamma N\Lambda}{2}.
	\end{eqnarray}
	We now turn to solve these equations of motion in the Painlev\'{e}-Gullstrand coordinates and set $N=1$. Taking into account the Hamiltonian constraint $\mathcal{H}^{(\mathrm{LQG})}_\Lambda$, the complete set of equations is as follows
	\begin{eqnarray}\label{pLQG}
		&\dot{E}^p=0;\quad \dot{p}=0\nonumber\\
		&N=1;\quad \mathcal{H}^{(\mathrm{LQG})}_\Lambda=0
	\end{eqnarray}
	Then by relating $\dot{E}^p=0$ and $N=1$, Eq. \eqref{EpLQG} implies three possibilities: \ba
1, \quad \partial_x{\left(\frac{E^p}x\right)}=0,\\
2, \quad \sin\frac{\sqrt\Delta p}x=0,\\
3, \quad \cos\frac{\sqrt\Delta p}x=0.
\ea For the case $1$ we assuming the term $\partial_x{\left(\frac{E^p}x\right)}$ in equation \eqref{EpLQG} vanishes, the solutions is 
	\begin{eqnarray}\label{solution of Ep}
		E^p=C_3x,
	\end{eqnarray} then, by using Eq. \eqref{pLQG}, we have
	\begin{eqnarray}
			\label{solution of p} p=\frac{x}{\sqrt{\Delta}}\arcsin\left[\left(\frac{1}{C^2_3}-1\right)\gamma^2\Delta\frac{1}{x^2}+\frac{C_4}{x^3}\right.\nonumber\\+\left.\frac{\gamma^2\Delta\Lambda}{3}\right]^{1/2},
	\end{eqnarray}
	where $C_3$ and $C_4$ are constants of integration. It is not hard to confirm that the solutions \eqref{solution of Ep} and \eqref{solution of p} satisfy the Hamiltonian constraint $\mathcal{H}^{(\mathrm{LQG})}_\Lambda=0$. For case $2$ and case $3$, use the same logic of \cite{Ewing 2020} and after careful calculations, it is determined that neither of them satisfies the Hamiltonian constraint. Therefore, the solution has only one form. 
    
    Next, we need to determine the values of these integration constants $C_3$ and $C_4$. Noticing equations \eqref{f square}, \eqref{y square}, \eqref{solution of Ep} and $N=1$, the metric \eqref{metric} can now be written in the following form 
    \begin{eqnarray}
    	ds^2=&-&\left(1-\left(C_3N^x\right)^2\right)dt^2+2C_3^2N^xdxdt\nonumber\\&+&C_3^2dx^2+x^2d\Omega^2,
    \end{eqnarray}
	where we should note that the LQG correction terms are present in $N^x$. Because we want the target metric to recover the classical case \eqref{PG metric} when the LQG corrections are removed, it is easy to know that $C_3^2=1$ in this case. Substituting $p$ \eqref{solution of p} into $N^x$ \eqref{NxLQG}, we obtain the desired result
	\begin{eqnarray}
		N^x=\left(\frac{x^2\Lambda}{3}+\frac{C_4}{\gamma^2\Delta x}\right)^{1/2}\left(1-\frac{\gamma^2\Delta\Lambda}{3}-\frac{C_4}{x^3}\right)^{1/2},
	\end{eqnarray} taking into account that $$\cos\frac{\sqrt\Delta p}x=-\sqrt{1-\sin^2\frac{\sqrt\Delta p}x}$$ for consistency with the classical case. When $\Delta\rightarrow0$, $N^x$ should revert to the classical case (which can be given by using equations \eqref{relation about N^x and N} and \eqref{classical solution of p}). Hence, we can ascertain that $C_4=2GM\gamma^2\Delta$. Then, the resulting LQG-corrected metric with the cosmological constant reads
	\begin{align}\label{LQG-corrected metric}
		 ds^2=-F(x)dt^2+2N^xdxdt+dx^2+x^2d\Omega^2,
	\end{align}
	with
	\begin{eqnarray}\label{Fx and Nx}
		F(x)=1-\left(N^x\right)^2,
	\end{eqnarray}
	\begin{eqnarray}\label{Nx}
		N^x=\left(\frac{x^2\Lambda}{3}+\frac{2GM}{ x}\right)^{1/2}\left(1-\frac{2GM\gamma^2\Delta}{x^3}\right.\nonumber\\-\left.\frac{\gamma^2\Delta\Lambda}{3}\right)^{1/2}.
	\end{eqnarray}
	From the above result, it can be easily verified that as $\Delta\rightarrow0$, the classical Schwarzschild de-Sitter(anti-de Sitter) solutions are recovered. Another point worth mentioning is that when $\Lambda$ is set to zero, the metric \eqref{LQG-corrected metric} can be restored to the case without cosmological constant as that in \cite{Ewing 2020}, which further demonstrates the correctness of the results of the present paper. Moreover, the quantum corrected metric obtained in current paper is coincide with the metric from the quantum Oppenheimer-Snyder collapse model \cite{Lewandowski 2023,Shao 2023}, which also add confidence on our model.

    \section{Physical properties of the effective metric}\label{Physicalpropertiesofmetric}
	\subsection{Geodesic and Effective Range}
	
	As stated in \cite{Ewing 2020} \cite{Lewandowski 2023}, the metric corrected by LQG is only valid for $x>x_{min}$. This is because, in the LQC, there is an upper limit to the energy density of any matter field. Therefore, in order to generate a gravitational field with mass $M$, the corresponding matter field cannot be infinitely small but will have a radius corresponding to the maximum energy density. Now we will search for the lower bound $x_{min}$. 
	
	Investigating geodesics provides a direct method for understanding the structure of spacetime, hence it is advisable to begin with them. To simplify the analysis, we will focus solely on radial geodesics that satisfy the following equation
	\begin{align}\label{geodesic equation}
		-\kappa=-F(x)\left(\frac{dt}{d\tau}\right)^2+2N^x\left(\frac{dx}{d\tau}\right)\left(\frac{dt}{d\tau}\right)+\left(\frac{dx}{d\tau}\right)^2
	\end{align}
	 where $\tau$ represents the geodesic parameter. When $\kappa=1$, it corresponds to timelike geodesics, and $\tau$ represents proper time. On the other hand, if $\kappa=0$, it refers to null geodesics, and $\tau$ represents a chosen affine parameter. It is important to note that in this case of a stationary spherically symmetric spacetime, $T^\mu=(\partial/\partial t)^\mu$ is a Killing vector. Setting the tangent vector of the geodesic as $\xi^\mu=(\partial/\partial \tau)^\mu$, then $T^\mu\xi_\mu$ is constant along the geodesic, which we refer to as the conserved energy written as 
	 \begin{eqnarray}
	 	E=T^\mu\xi_\mu=-F(x)\frac{dt}{d\tau}+N^x\frac{dx}{d\tau}.
	 \end{eqnarray}
	 
	 For timelike geodesics $(\kappa=1)$ and considering $E=1$, it corresponds to a stationary particle starting its motion from infinity. The geodesic equation \eqref{geodesic equation} can be further simplified as follows
	 \begin{eqnarray}\label{timelike geodesic}
	 	\frac{dx}{d\tau}=\pm \sqrt{1-F(x)},
	 \end{eqnarray}
	 where it can be observed that $x_{min}$ is at $dx/d\tau=0$ and the solution is 
	 \begin{eqnarray}\label{xmin}
	 	x_{min}=\left(\frac{6GM\gamma^2\Delta}{3-\gamma^2\Delta\Lambda}\right)^{1/3}.
	 \end{eqnarray}
	 Compared with the Schwarzschild case \cite{Ewing 2020}, it is easy to see that the term involving the cosmological constant causes a slight variation on $x_{min}$. However, when this term is neglected, we obtain exactly the same value of $x_{min}$ as discussed in \cite{Ewing 2020,Gambini 2020}. Moreover, since the $x_{min}$ should be greater than zero, then the denominator of Eq. \eqref{xmin} implies an upper bound on the cosmological constant as $\Lambda<\frac{3}{\gamma^2\Delta}$.
	 
	 Besides, for null geodesic, the situation will be much simpler. Since $\kappa=0$, equation \ref{geodesic equation} can be divided by $dt/d\tau$ and written as
	 \begin{eqnarray}\label{null geodesic}
	 	\frac{dx}{dt}=-N^x\pm1.
	 \end{eqnarray}
	 A nontrivial case arises when the outgoing null rays with $dx/dt=-N^x+1$ have zeros at $x>0$. However, the ingoing rays with $dx/dt=-N^x-1$ are always less than zero. One point worth mentioning here is that it is previously stated that $T^a$ represents a Killing vector field, and its corresponding Killing horizon is given by $T^aT_a=0$, which can be simplified as $F(x)=0$. But in fact, since $F(x)=0$ and $dx/dt=0$ are equivalent to each other, the zeros of $dx/dt$ also represent the Killing horizon. 
	 
\begin{figure}[h!]
	 	\centering
	 	\includegraphics[scale=1.05]{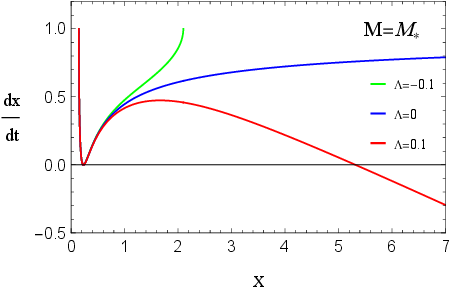}
	 	\caption{\label{fig null geodesic dx divided dt} For outgoing null geodesics, we have plotted the case of the black hole with mass $M_*$ and compared three different spacetimes. AdS spacetime is taken to have $\Lambda=-0.1$, while dS spacetime has $\Lambda=0.1$. It can be observed that the three spacetimes exhibit little difference near the Killing horizon, and only significant differences arise far from the horizon. The remaining parameters are taken as $\left\{\gamma=1, \Delta=0.01, G=1 \right\}$.}
	 \end{figure}

	 According to \cite{Ewing 2020}, $M_*=8\gamma\sqrt{\Delta}/\sqrt{27}G$ corresponding to the critical mass means that when $\Lambda=0$ and, there is only one Killing horizon. More precisely, as the mass is greater than $M_*$, there exist two Killing horizons, while if the mass is smaller than $M_*$, there are no Killing horizons. The detailed situation can be visualized in the figure \ref{fig null geodesic dx divided dt}. 
	 \begin{figure}
	 	\centering
	 	\includegraphics[scale=1.05]{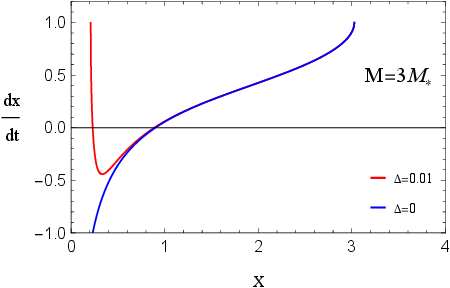}
	 	
	 	\includegraphics[scale=1.05]{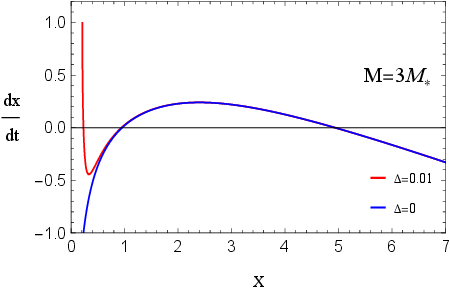}
	    \caption{\label{fig null geodesic ads and ds} For outgoing null geodesics, we have separately plotted the cases of the AdS black holes  (first figure) and the dS black holes (second figure) with the mass of $3M_*$. In both figures, the red curve represents the LQG-corrected metric, while the blue curve means the classical metric.  For AdS black holes, we consider $\Lambda=-0.1$, whereas for dS black holes, $\Lambda=0.1$. All the remaining parameters are taken as $\left\{\gamma=1, G=1 \right\}$.}
	 \end{figure}
	 
	 From Fig \ref{fig null geodesic ads and ds}, in the case of the LQG-corrected AdS black hole, the mass exceeds $M_*$, resulting in the presence of both inner and outer horizons. For LQG-corrected metric, the left endpoint of the red curve correspond to the minimum value of $x$, where $dx/dt=1$, as can be easily verified by equations \eqref{Fx and Nx}, \eqref{timelike geodesic}, and \eqref{null geodesic}. Conversely, the classical AdS black hole exhibits only one horizon, and $dx/dt$ diverges rapidly as $x \rightarrow 0$, indicating the presence of the spacetime singularity. In the case of dS black hole, the LQG correction also results in the existence of two Killing horizons, avoiding the singularity issue. Additionally, the cosmological horizon can be observed from the curves.
	 
	 \subsection{Curvature scalars and surface gravity}
	 Investigating the curvature of can provide us with a deeper understanding of the geometric structure of this spacetime. Furthermore, it allows for a more straightforward comparison to be made with the scenario where the cosmological constant is not present. With the corrected metric \ref{LQG-corrected metric}, various curvature scalars can be straightforwardly calculated. The results are summarized as follows:
	 \begin{eqnarray}
	 	R=4\Lambda-\frac{4}{3}\gamma^{2}\Delta\Lambda^{2}-\frac{24(G M\gamma)^{2}\Delta}{x^{6}},
	 \end{eqnarray}
	 \begin{eqnarray}
	 	R_{\mu\nu}R^{\mu\nu}=\frac49\Lambda^2\left(-3+\gamma^2\Delta\Lambda\right)^2+\frac{1440(GM\gamma)^4\Delta^2}{x^{12}}\nonumber\\+\frac{16(GM\gamma)^2\left(-3+\gamma^2\right)\Delta\Lambda}{x^6},
	 \end{eqnarray}
	 \begin{eqnarray}
	 	R_{\mu\nu\rho\sigma}R^{\mu\nu\rho\sigma}&=&\frac8{27}\Lambda^2\left(-3+\gamma^2\Delta\Lambda\right)^2\nonumber\\&+&\frac{320(GM)^3\gamma^2\left(-3+2\gamma^2\Delta\Lambda\right)\Delta}{x^9}\nonumber\\&+&\frac{32\left(GM\gamma\right)^2\left(-3+\gamma^2\Delta\Lambda\right)\Delta\Lambda}{x^6}\nonumber\\&+&\frac{7488(GM\gamma)^4\Delta^2}{{x}^{12}}.
	 \end{eqnarray}
	For physical reasons, the absolute value of curvature scalars should correspond to a maximum when $x$ approaches $x_{min}$ in vacuum. By substituting $x_{min}$ into various curvature scalars, we obtain the following results
	\begin{eqnarray}
		\lim_{x\to x_{\min}}R=-\frac6{\gamma^2\Delta}+8\Lambda-2\gamma^2\Delta\Lambda^2,
	\end{eqnarray}
	\begin{align}
		&\lim_{x\to x_{\min}}R_{\mu\nu}R^{\mu\nu}\nonumber\\&=\frac{2\left(-3+\gamma^2\Delta\Lambda\right)^2\left(5-4\gamma^2\Delta\Lambda+\gamma^4\Delta^2\Lambda^2\right)}{\gamma^4\Delta^2},
	\end{align}
	\begin{align}
		&\lim_{x\to x_{\min}}R_{\mu\nu\rho\sigma}R^{\mu\nu\rho\sigma}\nonumber\\
		&=\frac{4\left(-3+\gamma^2\Delta\Lambda\right)^2\left(10-6\gamma^2\Delta\Lambda+\gamma^4\Delta^2\Lambda^2\right)}{\gamma^4\Delta^2}.
	\end{align}
	These are also constants independent of the black hole mass. Unsurprisingly, when the terms containing the cosmological constant are discarded, these curvature scalars align with those in \cite{Ewing 2020} \cite{Gambini 2020}. 
	
\begin{figure}[h!]
		\centering
		\includegraphics[scale=1]{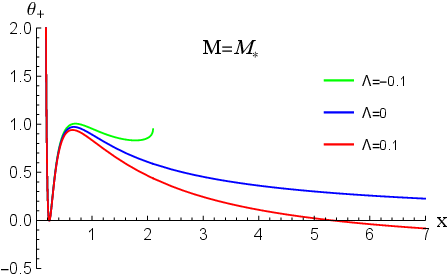}
		\caption{\label{fig expansion1}The expansions of the outgoing null geodesics are compared for different cosmological constants. In this case, the mass is chosen as $M_*$, indicating that when $\Lambda=0$ , there is only one event horizon. More detailed calculations reveal when $\Lambda=-0.1$, there is no event horizon and the curve's endpoint cannot extend indefinitely but reaches a certain point. This also implies the maximum value that $x$ can take. When $\Lambda=0.1$, there are two close inner and outer horizons, and a cosmological horizon can be clearly observed at the end of the curve. These properties are analogous to those of the outgoing null geodesics. The remaining parameters are taken as $\left\{\gamma=1, \Delta=0.01, G=1 \right\}$.}
		\includegraphics[scale=1]{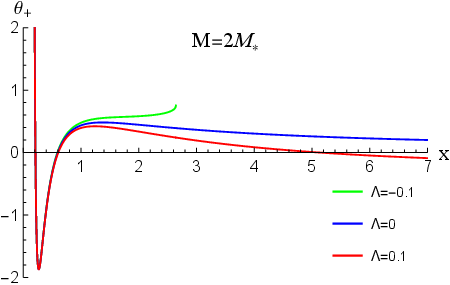}
		\caption{\label{fig expansion2} Comparison of the null expansion $\theta_+$ for the LQG-corrected metric with different values of the cosmological constant. The mass is $2M_*$, compared to the case with mass $M_*$, the most notable difference is that regardless of the value of the cosmological constant, the black hole exhibits both inner and outer event horizons. The influence of the cosmological constant is only evident at the endpoint of the curve. The remaining parameters are taken as $\left\{\gamma=1, \Delta=0.01, G=1 \right\}$.}
	\end{figure}

	In addition, exploring the trapped surface can further verify the results of the effective metric \ref{LQG-corrected metric} we have obtained. Considering the null geodesic congruence, due to the spherical symmetry, on a 2-dimensional sphere $\mathcal{S}$, the radial coordinate $x$ and the time coordinate $t$ are constant. Therefore, there are two independent null tangent vectors orthogonal to $\mathcal{S}$, referred to as outgoing null geodesics and ingoing null geodesics. These two geodesics correspond to the outgoing expansion parameter $\theta_+$ and ingoing expansion parameter $\theta_-$ respectively. By definition, the trapped surface corresponds to the region where both $\theta_+$ and $\theta_-$ are negative. Here, we define the tangent vectors of the outgoing null geodesics and the ingoing null geodesics to be the same as \cite{Ewing 2020}, denoted by 
	\begin{eqnarray}
		l^{\mu}=\left(\frac{\partial}{\partial t}\right)^\mu+(1-N^x)\left(\frac{\partial}{\partial x}\right)^\mu,
	\end{eqnarray}
	and
	\begin{eqnarray}
		k^{\mu}=\left(\frac{\partial}{\partial t}\right)^\mu+(-1-N^x)\left(\frac{\partial}{\partial x}\right)^\mu.
	\end{eqnarray}
    Normalize these two vector fields such that $l^\mu k_\mu=-2$ and the hypersurface metric for $\mathcal{S}$ is 
	\begin{eqnarray}
	h_{\mu\nu}=g_{\mu\nu}+\frac{1}{2}\left(l_\mu k\nu+k_\mu l_\nu\right).
	\end{eqnarray}
	Based on the definition of the expansions $\theta_+$ and $\theta_-$, the simple expression can be obtained through straightforward calculations and the results are 
	\begin{eqnarray}\label{expansion}
		\theta_+&=&h^{\mu\nu}\nabla_\mu l_\nu=\frac{2}{x}\left(1-N^x\right),\\
		\theta_-&=&h^{\mu\nu}\nabla_\mu k_\nu=-\frac{2}{x}\left(1+N^x\right).
	\end{eqnarray} 
	Here, $\theta_-$ remains negative within the range of $x$, so considering $\theta_+$ alone is sufficient. The main results are displayed in Figure \ref{fig expansion1} and \ref{fig expansion2}. Besides, the trapped surface tends to increase with the increase of mass. From the expression of the expansion \eqref{expansion} and the null geodesic \eqref{null geodesic}, it can be observed that they differ only by a factor of $2/x$. Therefore, it is conceivable that the zero points of the null geodesics align with the outgoing expansion and the physical interpretation is consistent with the null geodesics.
	
\begin{figure}
		\centering
		\includegraphics[scale=1]{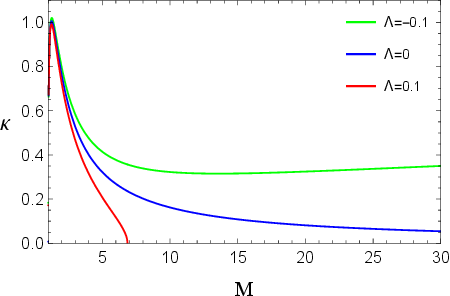}
		\caption{\label{fig surface gravity} By taking the mass of the black hole as the independent variable, we compared the surface gravity under different values of the cosmological constant. The mass is measured in units of $M_*$. The minimum value of the mass is set at $M_*$ because when the mass is less than $M_*$, there is no horizon, and any discussion concerning surface gravity becomes meaningless. The remaining parameters are taken as $\left\{\gamma=1, \Delta=0.01, G=1 \right\}$.}
	\end{figure}

	When a black hole possesses an outer horizon, it becomes straightforward to calculate the surface gravity by employing 
	\begin{eqnarray}
		\kappa=\left.\frac{1}{2}\frac{\partial F(x)}{\partial x}\right|_{x=x_{h}}.
	\end{eqnarray} 
	where $x=x_{h}$ represents the position of the outer horizon. Due to the complexity of the expression of $F(x)$, it is not practical to calculate the analytical solution of the surface gravity directly. Moreover, if we adopt a first-order approximation for $\Delta$, the distinctions between the three types of spacetime will not be apparent and the result of the expression will be the same. Therefore, we adopt a numerical approach to solve the problem. The detailed results are shown in Figure \ref{fig surface gravity}. 
	The green curve corresponds to the AdS black hole, and the red curve represents the dS black hole. At smaller masses, due to the LQG correction, they behave almost identically and both exhibit a peak. However, as the mass increases, due to the distinct characteristics of the spacetime, the surface gravity of the dS black hole rapidly tends to zero and eventually reaches an upper mass limit, whereas the surface gravity of the AdS black hole first decreases and then slowly increases.
	  
	 \section{Thermodynamics}\label{Thermodynamic LQG}
\begin{figure}[h!]
      	\centering
      	\includegraphics[scale=1.1]{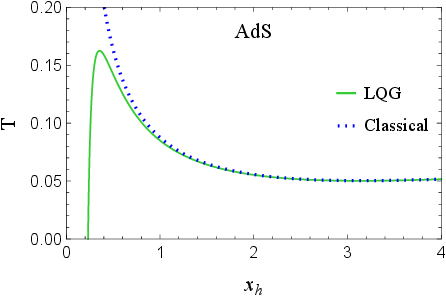}
      	\caption{\label{fig T-rh} $T-x_h$ diagram of two types of AdS black holes. The green solid line represents the LQG case, while the blue dashed line signifies the classical case. It indicates that when the radius is extremely small, the temperature of the LQG black hole decreases as the mass increases. The parameters are taken as $\left\{\gamma=1, \Delta=0.01, \Lambda=-0.1\right\}$.}
      \end{figure}

\begin{figure}[h!]
      	\centering
      	\includegraphics[scale=1.1]{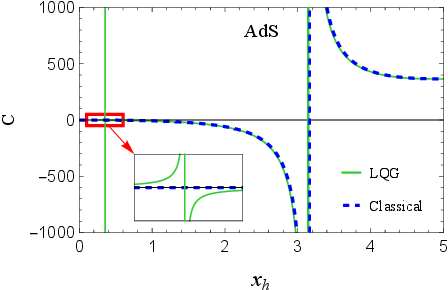}
      	\caption{\label{fig C-rh} $C-x_h$ diagram of two types of AdS black holes. The green vertical lines $\left(x\approx0.38,x\approx3.18\right)$ mark the heat capacity divergence for LQG black holes, and the vertical blue dashed line $\left(x\approx3.18\right)$ does the same for classical black holes. The left green vertical line $\left(x\approx0.38\right)$, being so close to the heat capacity curve of the LQG black hole that they appear as a single vertical line. Both LQG and classical black holes exhibit a divergence in heat capacity near $x\approx3.18$, suggesting the occurrence of a phase transition. This transition signifies a shift in thermodynamic stability from instability $(C<0)$ to stability $(C>0)$. Additionally, the magnified view in the figure reveals that the LQG black hole exhibits an extra phase transition compared to the classical black hole. At smaller horizon radius (x<0.38), contrary to the classical case, the LQG black hole is stable. The parameters are taken as $\left\{\gamma=1, \Delta=0.01, \Lambda=-0.1 \right\}$.}
      \end{figure}

	   Within this subsection, it is  specified that we are adopting the natural units exclusively $( G = \hbar = c = k_B = 1)$. From equations \ref{Fx and Nx} and \ref{Nx}, the mass of black holes can be expressed in terms of the outer horizon $x_{h}$,
	  \begin{eqnarray}\label{thermodynamic M}
	  	M=\frac{x_{h}^{3}\left(3-2\gamma^{2}\Delta\Lambda\right)-3x_{h}^{2}\sqrt{x_{h}^{2}-4\gamma^{2}\Delta}}{12\gamma^{2}\Delta}.
	  \end{eqnarray}
      The Hawking temperature of the black holes can be easily obtained by requiring the absence of conical singularity at the horizon in the Euclidean sector of the black hole solution,
      \begin{eqnarray}\label{thermodynamic T}
      	T&=&\frac{1}{4 \pi}\left.\frac{\partial F(x)}{\partial x}\right|_{x=x_h}\nonumber\\
      	&=&\frac{-3x_{h}^{2}+x_{h}\sqrt{x_{h}^{2}-4\gamma^{2}\Delta}\left(3-2\gamma^{2}\Delta\Lambda\right)+8\gamma^{2}\Delta}{8\pi x_{h}\gamma^{2}\Delta}\nonumber,\\
      \end{eqnarray}
      and the corresponding figure \ref{fig T-rh} is plotted. Interestingly, we observed that for a small black hole, the temperature of the LQG black hole decreases as the mass decreases, which is quite different with the classical scenario. However, in some modified gravity theories, the corresponding Hawking temperature also exhibits the similar behavior \cite{Hongshengzhang23}.
      The heat capacity of black holes reads
      \begin{eqnarray}
      	C=\left(\frac{\partial M}{\partial x_h}\right)/ \left(\frac{\partial T}{\partial x_h}\right),
      \end{eqnarray}
       and the corresponding figure \ref{fig C-rh} is also plotted. The figure \ref{fig C-rh} reveals that the LQG black hole exhibits an extra phase transition compared to the classical black hole.
      
      Classical black holes generally satisfy the so-called area law, which states that the entropy of a black hole is equal to one-quarter of the horizon area. However, in the presence of quantum corrections, the area-entropy law no longer holds. In recent years, research within the framework of LQG has shown that black hole entropy, compared to the classical result, includes an additional logarithmic correction term \cite{Frodden 2012}. We now assume that the thermodynamic first law in the extended phase space holds for our model, and thus we have \cite{Kubiznak 2012}:
     \begin{eqnarray}
      	dM=TdS+VdP.
     \end{eqnarray}
	 The entropy of a black hole is
	 \begin{eqnarray}
	 	S&=&\int \left(T^{-1}dM\right)_P=\int_{x_t}^{x_h}\left[T^{-1}\left(\frac{\partial M}{\partial x_h}\right)\right]_P dx_h \nonumber\\
	 	&=&\pi x_h^2\sqrt{1-\frac{4\gamma^2\Delta}{x_h^2}}\nonumber\\
	 	&-&2\pi\gamma^{2}\Delta \ln\left(-x_h+x_h\sqrt{1-\frac{4\gamma^2\Delta}{x_h^2}}\right)^{2}+S_0\nonumber\\
	 	&=&\frac{A}{4}+2\pi\gamma^2\Delta\ln\left(\frac{A}{16\pi\gamma^4\Delta^2}\right)+S_1+O\left[\frac1{x_h}\right],
	 \end{eqnarray}
      where $S_0$, $S_1$ are constants and $A=4\pi x_h^2$. Since the entropy should approach zero as temperature nears zero, we set the lower limit of the integral to  $x_t$, which represents the horizon radius of the black hole when the temperature vanishes. The final result employs an approximation that $x_h$ is much greater than $4\gamma^{2}\Delta$, so the square root term can be rewritten as 
     \begin{eqnarray}
      \sqrt{1-\frac{4\gamma^2\Delta}{x_h^2}}=1-\frac{2\gamma^2\Delta}{x_h^2}+O\left[\frac1{x_h}\right]^3.
     \end{eqnarray}
     Therefore, the final expression for the entropy can be approximately considered as the area law plus a quantum logarithmic correction term. 
     
\begin{figure}[h!]
     	\centering
     	\includegraphics[scale=1.1]{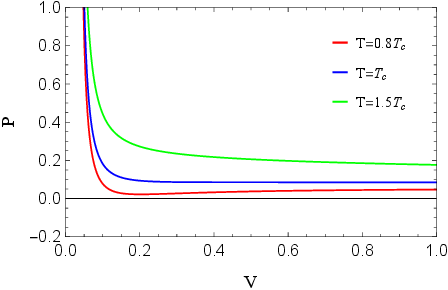}
     	\caption{\label{fig P} Graphic representation of the state equation with the values $\left\{\gamma=1, \Delta=0.01, \Lambda=-0.1 \right\}$.}
     \end{figure}

\begin{figure}[h!]
     	\centering
     	\includegraphics[scale=1.1]{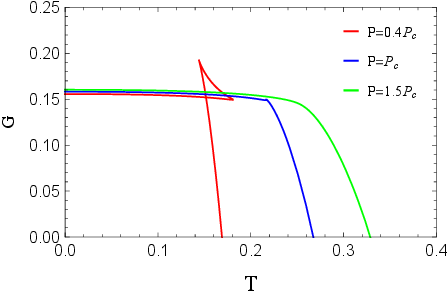}
     	\caption{\label{fig G} This picture shows the characteristic swallowtail behaviour of the Gibbs free energy as a function of temperature. It is observed that when $P < P_c$, a first-order phase transition occurs, and at each temperature, there can be up to three values of free energy, each representing a different phase of the black hole.}
     \end{figure}

     In AdS black holes, the variation of $\Lambda$ can be incorporated into the first law of thermodynamics, and it is no longer treated as a constant. The pressure is understood to be a quantity related to $\Lambda$ \cite{Kastor 2009}:
     \begin{eqnarray}\label{thermodynamic P}
     	P=-\frac{\Lambda}{8\pi}.
     \end{eqnarray}
     Therefore, $\Lambda$ in the equation \ref{thermodynamic M} can be expressed as $-8\pi P$, and by calculating the partial derivative $\partial M/\partial P$, the thermodynamic volume of the black hole can be obtained by,
     \begin{eqnarray}\label{thermodynamic V}
     	V=\frac{4\pi x_h^3}{3}.
     \end{eqnarray}
      Utilizing equations \ref{thermodynamic P} and \ref{thermodynamic V}, the temperature \ref{thermodynamic T} can be expressed in terms of variables $P$ and $V$, and after rearrangement, the equation of state can be obtained by,
     \begin{eqnarray}\label{PofTV}
       P(T,V)=\frac{Q_1+Q_2}{Q_3},
     \end{eqnarray}
     \begin{eqnarray}
       Q_1&=&9\sqrt[3]{6V^2}-16\sqrt[3]{(6\pi)}\gamma^2\Delta,\\
       Q_2&=&3\sqrt[3]{V}\left(16\pi^{4/3}\gamma^2\Delta T-3\sqrt{(6V)^{2/3}-16\pi^{2/3}\gamma^2\Delta}\right), \nonumber \\
       \\
       Q_3&=&48\pi\gamma^2\Delta\sqrt[3]{V}\sqrt{(6V)^{2/3}-16\pi^{2/3}\gamma^2\Delta}.
     \end{eqnarray}
     We can search for the critical point through the following conditions,
     \begin{eqnarray}
     	\left(\frac{\partial P}{\partial V}\right)_T=0,\text{ and }\left(\frac{\partial^2P}{\partial V^2}\right)_T=0.
     \end{eqnarray}
     The critical values of temperature $T_c$, volume $V_c$ and pressure $P_c$ are given, respectively, by
     \begin{eqnarray}
     	T_c&=&\frac1{6\pi\gamma\sqrt{6 \Delta}},\\
     	V_c&=&64\sqrt{6}\pi \gamma^3 \sqrt{\Delta^3},\\
     	P_c&=&\frac{-27+5\sqrt{30}}{144\pi\gamma^2\Delta},
     \end{eqnarray}
     and we show figure \ref{fig P} representing pressure as a function of volume. The solution has isotherms similar to the Van der Walls fluids.
     Having the equation of state $P(T,V)$, we can define the compressibility factor,
     \begin{eqnarray}
     	Z\left(T,V\right)=\frac{PV}{T},
     \end{eqnarray}
     which indicates how much a gas deviates from ideal gases. When $Z = 1$, it implies that there is no interaction between particles. At the critical point, 
     \begin{eqnarray}
     	Z_c=\frac{P_c V_c}{T_c}=16\left(-27+5\sqrt{30}\right)\pi\gamma^2\Delta.
     \end{eqnarray}
     which indicates that the value of $Z_c$ is dependent on $\gamma$. Moreover, in the extended phase space, we consider that the function $M$ play the role of enthalpy $H$. So, the Gibbs free energy is written as
     \begin{eqnarray}
     	G=M-TS,
     \end{eqnarray}
      and the behaviour of $G$ is shown in figure \ref{fig G}.
 
     Critical exponents describe the behaviour of physical quantities near the critical point. It is believed that they are universal. To obtain them, we use the so-called reduced variables
     \begin{eqnarray}\label{reduced variables}
     	\tilde{t}=\frac{T-T_c}{T_c},\quad \tilde{v}=\frac{V-V_c}{V_c}, \quad\tilde{p}=\frac{P}{P_c}.
     \end{eqnarray} 
     We define the critical exponents as $\alpha, \beta, \lambda$ and $\chi$. They are specified by the following equations,
     \begin{eqnarray}
      	C_{V}=T\left(\frac{\partial S}{\partial T}\right)\propto\left|\tilde{t}\right|^{-\alpha}, \\
      	\eta=V_{1}-V_{2}\propto\left|\tilde{t}\right|^{\beta}, \\
      	k_{T}=-\frac{1}{V}\frac{\partial V}{\partial P}\propto\left|\tilde{t}\right|^{-\lambda}, \label{kT}\\
      	\left|P-P_c\right|\propto\left|V-V_c\right|^\chi,
      \end{eqnarray}
      where $C_{V}$, $\eta$, $k_T$ are respectively the heat capacity at constant volume, the difference in volume between two phases and the isothermal compressibility. Clearly, the entropy $S$ is a function that does not explicitly depend on $T$, so it can be determined that $\alpha=0$. If we rewrite the equation of state \ref{PofTV} by using eq.\ref{reduced variables} and expand it for small values of $\tilde{t}$ and $\tilde{v}$, we get 
      \begin{eqnarray}\label{tilde p}
      	\tilde{p}=1+\left(B_1-B_2\tilde{v}\right)\tilde{t}-B_3\tilde{v}^3+O\left(\tilde{v}^4,\tilde{t}\tilde{v}^2\right),
      \end{eqnarray}
     where $B_1$, $B_2$ and $B_3$ are positive constants. If we consider $\tilde{t}$ as constant and derive $\tilde{p}$ to $\tilde{v}$, we obtain
     \begin{eqnarray}
     	d\tilde{p}=-\left(B_2\tilde{t}+3B_3\tilde{v}^2\right)d\tilde{v}.
     	\end{eqnarray}
     This relation is important to apply Maxwell’s area law, which states \cite{Dayyani 2018}, \cite{Wei 2020}
     \begin{eqnarray}
     	\int VdP=0.
     \end{eqnarray}
     Thus, when considering Maxwell’s area law and the fact that the pressure is constant at the phase transition, we get
     \begin{eqnarray}
     \tilde{p} &=& 1+(B_1-B_2\tilde{v}_1)\tilde{t}-B_3\tilde{v}_1^3\nonumber\\
     &=&1+(B_1-B_2\tilde{v}_2)\tilde{t}-B_3\tilde{v}_2^3,\\
     0&=& \int_{\tilde{v}_1}^{\tilde{v}_2}(\tilde{v}+1)\left(B_2\tilde{t}+3B_3\tilde{v}^2\right)d\tilde{v}.
     \end{eqnarray}
     Solving these equations yields the nontrivial solution as follows:
     \begin{eqnarray}
     \tilde{v}_1=-\tilde{v}_2=\frac{B_2}{B_3}\sqrt{-\tilde{t}}\quad  \text{and thus} \quad \eta\propto\sqrt{-\tilde{t}},
     \end{eqnarray}
     which means that one of the critical exponents is $\beta = 1/2$. We can also use equations \ref{kT} and \ref{tilde p} to calculate the isothermal compressibility and we find that
     \begin{eqnarray}
     k_T\propto1/ \tilde{t},
     \end{eqnarray}
     such that $\lambda=1$. To obtain the final critical exponents, we rewrite the condition $\left|P-P_c\right|\propto\left|V-V_c\right|^\chi\mathrm{~as~}\left|\tilde{p}-1\right|\propto\left|\tilde{v}\right|^\chi$. From equation \ref{tilde p}, we get
     \begin{eqnarray}
     \tilde{p}-1\propto (B_1-B_2\tilde{v})\tilde{t}-B_3\tilde{v}^3.
     \end{eqnarray}
     It is worth reminding that in the critical isotherm we have $\tilde{t} = 0$, so that,
     \begin{eqnarray}
     \tilde{p}-1\propto \tilde{v}^3.
     \end{eqnarray}
     In general, the critical exponents are $\alpha=0, \beta=1/2, \lambda=1,$ and $\chi=3$.
	 \section{Conclusions and discussions}\label{Sec-Conclusions and discussions}
	 The main result in this paper is obtained based on the Painlev\'{e}-Gullstrand coordinate system and utilizes the components of the Ashtekar-Barbero connection and density-triad to describe a 4-dimensional spherically symmetric spacetime with a cosmological constant. We first perform calculations in the classical scenario. Starting from the gravitational action, after symmetry reduction, we consider the constrained system to consist of Hamiltonian constraint (scalar constraint) and the diffeomorphism constraint (vector constraint) due to the constrained nature of the system. By exploiting the spherical symmetry, further simplification of the constraints can be achieved. The equations of motion can then be obtained.
	 
	 Then we impose a gauge fixing on the area. As a result, the number of dynamical variables is reduced from four to two, and the equations of motion are also reduced to two. By directly solving the equations, the classical Schwarzschild de-Sitter(anti-de Sitter) solutions are obtained as expected.
	
	Next, we proceed to obtain the LQG effective dynamics for vacuum spherically symmetric space-times. We modified the variable $p$ of the connection by incorporating holonomy corrections and adjusted the Hamiltonian constraint, symplectic structure, and shift vector accordingly. Similarly, by solving the modified effective equations of motion, we obtained the LQG-corrected spherically symmetric metric with a cosmological constant. In comparison with the previous treatments of LQG black hole models, our treatment leads to a signiﬁcant difference. In \cite{Sartini 2021}, the authors investigated the quantum dynamics of the black hole interior with the cosmological constant in the unimodular gravity framework. While in our current paper, our start point is GR plus cosmological constant, we do not view the cosmological constant as a dynamical variable. As a result, analytical solutions for Schwarzschild de-Sitter(anti-de Sitter) black hole with LQG correction are obtained. 
	
	Due to the LQG correction, the variable $x$ has a restricted range of values. For AdS and dS spacetimes, we calculated their common minimum value $x_{min}$ using timelike geodesics. The dS spacetime exhibits the expected cosmological horizon. Compared to $\Lambda=0$, the effect of the cosmological constant is not only noticeable at distances far from the event horizon but also alter the geometry near the singularity. Moreover, there exist an upper bound on the cosmological constant as $\Lambda<\frac{3}{\gamma^2\Delta}$ obtained by Eq.\eqref{xmin}. 
	
	We also investigated the curvature scalars $R, R_{\mu\nu}R^{\mu\nu}, R_{\mu\nu\rho\sigma}R^{\mu\nu\rho\sigma}$ of the LQG-corrected metric, which are bounded at $x_{min}$ due to quantum gravity effects and are constants that only depend on $\Lambda$ and $\Delta$. Moreover, when setting $\Lambda=0$, all the curvature scalars can be recovered in the form presented in article \cite{Ewing 2020}. 
	
	It is worth mentioning that the computation of $x_{min}$ can also be directly obtained from $F(x)$. Since $F(x)=1-\left(N^x\right)^2$, it can be observed that $F(x)\leq 1$, and when $F(x) = 1$, it represents the boundary of applicability for $F(x)$. And $x_{min}$ is one of the solutions to $F(x) = 1$.

	
	In addition, we also calculate the thermodynamic properties of AdS black holes in LQG. The results indicate that the modification introduced by LQG do not lead to temperature divergence. For a small black hole, the temperature of the LQG black hole decreases as the mass decreases. Moreover, the LQG corrections also introduce an extra phase transition in the black hole's heat capacity at smaller radius, suggesting that the thermodynamic properties of small black holes in LQG might be stable. Then, we calculate the entropy of the AdS black holes in LQG, the result shows that a logarithmic term appeared as the leading order correction to the Beikenstein-Hawking entropy. Under the premise of the extended first law of thermodynamics being valid, the $P-V$ diagram shows that AdS black holes in LQG, like RN-AdS black holes \cite{Kubiznak 2012}, undergo phase transitions. We further calculated the critical exponents, and the results are consistent with those of the Van der Walls fluid. Last but not least, for the general covariance issue, for Schwarzschild case, the quantum framework of \cite{Ewing 2020} is equivalent to that in \cite{Gambini 2020}, and the latter is showing to be general covariance \cite{Pullin22a}. The framework of \cite{Pullin22a} can be generalized to include a cosmological constant in principle. However, since even in Schwarzschild case, the general covariance issue is still under debating \cite{Bojowald22,Pullin22b}, we would like to leave this topic for future study.
	
\begin{acknowledgements}
This work is supported by National Natural Science Foundation of China (NSFC) with Grants No.12275087. 
\end{acknowledgements}	



\begin{thebibliography}{10}
		 
		 
		 \bibitem{Ashtekar 2004}
		 A. Ashtekar and J. Lewandowski, Background Independent Quantum Gravity: A Status Report, Class. Quantum Grav. 21, R53 (2004).
		 
		 \bibitem{Rovelli 2004}
		 C. Rovelli, \textit{Quantum Gravity} (Cambridge University Press, Cambridge, England, 2004).
		 
		 \bibitem{Thiemann}
		 T. Thiemann, \textit{Modern Canonical Quantum General Relativity} (Cambridge University Press, Cambridge, United Kingdom, 2007).
		 		 		
		 \bibitem{Han 2007}
		 M. Han, Y. Ma, and W. Huang, Fundamental structure of loop quantum gravity, Int. J. Mod. Phys. D 16, 1397 (2007).
		 
		 \bibitem{Rovelli 1995}
		 C. Rovelli and L. Smolin, Discreteness of Area and Volume in Quantum Gravity, Nuclear Physics B 442, 593 (1995).
		 
		 \bibitem{Ashtekar 1997-01}
		 A. Ashtekar and J. Lewandowski, Quantum Theory of Geometry: I. Area Operators, Class. Quantum Grav. 14, A55 (1997).
		 
		 \bibitem{Ashtekar 1997-11}
		 A. Ashtekar and J. Lewandowski, Quantum Theory of Geometry II: Volume Operators, Adv. Theor. Math. Phys. 1, 388 (1997).
		 
		 \bibitem{Yang 2016}
		 J. Yang and Y. Ma, New Volume and Inverse Volume Operators for Loop Quantum Gravity, Phys. Rev. D 94, 044003 (2016).
		 
		 \bibitem{Thiemann 1998}
		 T. Thiemann, A Length Operator for Canonical Quantum Gravity, Journal of Mathematical Physics 39, 3372 (1998).
		 
		 \bibitem{Ma 2010}
		 Y. Ma, C. Soo, and J. Yang, New Length Operator for Loop Quantum Gravity, Phys. Rev. D 81, 124026 (2010).
		 		 		 
		 \bibitem{Zhang 2011}
		 X. Zhang and Y. Ma, Extension of Loop Quantum Gravity to f (R) Theories, Phys. Rev. Lett. 106, 171301 (2011).
		 
		 \bibitem{Bodendorfer 2013}
		 N. Bodendorfer, T. Thiemann, and A. Thurn, New Variables for Classical and Quantum Gravity in All Dimensions I. Hamiltonian Analysis, Class. Quantum Grav. 30, 045001 (2013).
		 
		 \bibitem{Zhang 2020}
		 X. Zhang, J. Yang, and Y. Ma, Canonical Loop Quantization of the Lowest-Order Projectable Horava Gravity, Phys. Rev. D 102, 124060 (2020).	 
 
		 \bibitem{Tan 2006}
		 H. Tan and Y. Ma, Semiclassical States in Homogeneous Loop Quantum Cosmology, Class. Quantum Grav. 23, 6793 (2006).
		 		 
		 \bibitem{Bojowald 2001}
		 M. Bojowald, Absence of a Singularity in Loop Quantum Cosmology, Phys. Rev. Lett. 86, 5227 (2001).
		 
		 \bibitem{Modesto 2004}
		 L. Modesto, Disappearance of the Black Hole Singularity in Loop Quantum Gravity, Phys. Rev. D 70, 124009 (2004).
		 
		 \bibitem{Ashtekar and Bojowald 2005}
		 A. Ashtekar and M. Bojowald, Quantum Geometry and the Schwarzschild Singularity, Class. Quantum Grav. 23, 391 (2005).	
		 
		 \bibitem{Ashtekar 2003}
		 A. Ashtekar, M. Bojowald, and J. Lewandowski, Mathematical Structure of Loop Quantum Cosmology, Advances in Theoretical and Mathematical Physics 7, 233 (2003).
		 
		 \bibitem{Bojowald 2008}
		 M. Bojowald,  Loop Quantum Cosmology, Living Rev. Relativ. 11, 4 (2008).
		 
		 \bibitem{Ashtekar 2011}
		 A. Ashtekar and P. Singh, Loop Quantum Cosmology: A Status Report, Class. Quantum Grav. 28, 213001 (2011).
		 		 
		 \bibitem{Boehmer 2007}
		 C. G. Boehmer and K. Vandersloot, Loop Quantum Dynamics of the Schwarzschild Interior, Phys. Rev. D 76, 104030 (2007).
		 
		 \bibitem{D.-W. 2008}
		 D.-W. Chiou, Phenomenological Loop Quantum Geometry of the Schwarzschild Black Hole, Phys. Rev. D 78, 064040 (2008).
		 
		 \bibitem{Ashtekar 2018}
		 A. Ashtekar, J. Olmedo, and P. Singh, Quantum Transfiguration of Kruskal Black Holes, Phys. Rev. Lett. 121, 241301 (2018).

         \bibitem{Zhang2020}C. Zhang and X. Zhang, Quantum geometry and effective dynamics of Janis-Newman-Winicour singularities, Phys. Rev. D 101, 086002 (2020),
		 
		 \bibitem{Zhang 2023}
		 X. Zhang, Loop Quantum Black Hole, Universe 9, 313 (2023).
		 
		 \bibitem{Sartini 2021}F. Sartini and M. Geiller, Quantum Dynamics of the Black Hole Interior in LQC, Phys. Rev. D 103, 066014 (2021).
		  \bibitem{Ewing 2020}
		 J. G. Kelly, R. Santacruz, and E. Wilson-Ewing, Effective Loop Quantum Gravity Framework for Vacuum Spherically Symmetric Spacetimes, Phys. Rev. D 102, 106024 (2020).		 		 
		 
         \bibitem{Friemann08}J. Friemann, M. Turner, D. Huterer, Dark energy and the accelerating Universe, Ann. Rev. Astron. Astrophys. 46, 385 (2008).

         \bibitem{Banerjee} N. Benerjee and D. Pavon, Cosmic acceleration without quintessence, Phys. Rev. D 63, 043504
         (2001).

          \bibitem{Sen}S. Sen and A. A. Sen, Late time acceleration in Brans-Dicke cosmology, Phys. Rev. D 63, 124006(2001).

          \bibitem{Qiang}L. Qiang, Y. Ma, M. Han and D. Yu,  5-dimensional Brans-Dicke theory and cosmic acceleration, Phys. Rev. D 71,  061501(R) (2005).

          \bibitem{Peebles03}P. J. E. Peebles and Bharat Ratra, The cosmological constant and dark energy, Rev. Mod. Phys. 75, 559 (2003).

          \bibitem{Tye01}S.-H. Henry Tye and Ira Wasserman, Brane World Solution to the Cosmological Constant Problem, Phys. Rev. Lett. 86, 1682 (2001).

          \bibitem{Weinberg89}S. Weinberg, The cosmological constant problem, Rev. Mod. Phys. 61, 1 (1989).

          \bibitem{Ashtekar16}A. Ashtekar, B. Bonga, and A. Kesavan, Gravitational Waves from Isolated Systems Surprising Consequences of a Positive Cosmological Constant, Phys. Rev. Lett. 116, 051101 (2016).
       
		 \bibitem{Bardeen 1973}J. M. Bardeen, B. Carter, and S. W. Hawking, The Four Laws of Black Hole Mechanics, Commun.Math. Phys. 31, 161 (1973).
		 		 
		 \bibitem{Wald 2001}R. M. Wald, The Thermodynamics of Black Holes, Living Reviews in Relativity 4,6 (2001).
		 		 
		 \bibitem{Shao 2023}
		 C.-Y. Shao, C. Zhang, W. Zhang, and C.-G. Shao, Scalar fields around a loop quantum gravity black hole in de Sitter spacetime: Quasinormal modes, late-time tails and strong cosmic censorship, Phys. Rev. D 109, 064012 (2024).
		 
		 \bibitem{Kubiznak 2012}
		 D. Kubiznak and R. B. Mann, P-V Criticality of Charged AdS Black Holes, J. High Energ. Phys. 2012, 33 (2012).		 
		 
		 \bibitem{Dayyani 2018}Z. Dayyani, A. Sheykhi, M. H. Dehghani, and S. Hajkhalili, Critical Behavior and Phase Transition of Dilaton Black Holes with Nonlinear Electrodynamics, Eur. Phys. J. C 78, 1 (2018).
		 
		 
		 \bibitem{Wei 2020}S.-W. Wei and Y.-X. Liu, Extended Thermodynamics and Microstructures of Four-Dimensional Charged Gauss-Bonnet Black Hole in AdS Space, Phys. Rev. D 101, 104018 (2020).

         \bibitem{Shao 2023}
		 C.-Y. Shao, C. Zhang, W. Zhang, and C.-G. Shao, Scalar fields around a loop quantum gravity black hole in de Sitter spacetime: Quasinormal modes, late-time tails and strong cosmic censorship, Phys. Rev. D 109, 064012 (2024).
			 
	     \bibitem{BV07}C. G. Böhmer and K. Vandersloot, Loop quantum dynamics of the Schwarzschild interior, Phys. Rev. D 76, 104030 (2007). 

         \bibitem{Hongshengzhang23}Y. Huang, D. Liu and H. Zhang, Novel black holes in higher derivative gravity, J. High Energ. Phys. 2023, 57 (2023).

		 \bibitem{Gambini 2020}
		 R. Gambini, J. Olmedo, and J. Pullin, Spherically symmetric loop quantum gravity: Analysis of improved dynamics, Classical Quantum Gravity 37, 205012 (2020).
		 
		 \bibitem{Lewandowski 2023}
		 J. Lewandowski, Y. Ma, J. Yang, and C. Zhang, Quantum Oppenheimer-Snyder and Swiss Cheese Models, Phys. Rev. Lett. 130, 101501 (2023).
		 
		 \bibitem{Frodden 2012}E. Frodden, A. Ghosh, and A. Perez, Black Hole Entropy in LQG: Recent Developments, in (Madrid, Spain, 2012), pp. 100–115.
		 
		 
         \bibitem{Kastor 2009}D. Kastor, S. Ray, and J. Traschen, Enthalpy and the Mechanics of AdS Black Holes, Class. Quantum Grav. 26, 195011 (2009).
         
         \bibitem{Pullin22a}R. Gambini, J. Olmedo, and J. Pullin, Towards a quantum notion of covariance in spherically symmetric loop quantum gravity, Phys. Rev. D 105, 026017 (2022).
        
         \bibitem{Bojowald22}M. Bojowald, Comment on "Towards a quantum notion of covariance in spherically symmetric loop quantum gravity" Phys. Rev. D 105, 108901 (2022)
        
         \bibitem{Pullin22b}R. Gambini, J. Olmedo, and J. Pullin, Reply to "Comment on ‘Towards a quantum notion of covariance in spherically symmetric loop quantum gravity"’, Phys. Rev. D 105, 108902 (2022).
         
         
         
         
	\end{thebibliography}
\end{document}